\documentclass[a4paper,12pt]{article}
\usepackage[]{graphicx}
\usepackage{amsmath}
\usepackage{amsfonts}
\usepackage{amssymb}

\title{On some fundamental problems of the theory of gravitation }
\author{L. V. Verozub, Kharkov National University}
\date{}
\begin{document}

\maketitle

\begin{abstract}
Cosmological observations indicate that the Einstein equation may not be entirely correct 
to describe gravity.
 However, numerous modifications of these equations usually do not affect foundations of 
the theory. 
In this paper  two important issue that lead to a substantial revision  of the theory are 
considered : \\
1. The significance of relativity of space-time geometry with respect to measuring instruments 
for theory of gravitation.\\
2. The gauge transformations of  the field variables in correct  theory of  gravitation.

\end{abstract}

\section{Relativity of Space-Time}

Einstein's theory of gravity is a realization of the idea of the relativity of the 
properties of space-time with respect to the distribution of matter. However, before the 
advent 
of Einstein's theory, Henri Poincar\'{e} showed that the properties of space and 
time are also 
relative to the properties of the used measuring instruments.
Of course now it can be said also about the properties of space-time too. However, 
these convincing arguments have never
 been implemented in physical theory. 

We can make a step towards the realization of this idea, if we will pay attention 
 that the properties of measuring instruments are one of the characteristics of 
the reference frame used.
 We can, therefore, expect that  we deal with the 
manifestation 
of a fundamental property of physical reality --- with 
space-time relativity with respect to the reference frame used.

The following simple example shows that this rather unexpected statement makes 
sense. 
Consider two reference frames, and two observers which proceed from the notion 
 of the relativity of space-time in the sens of 
Berkley-Leibnitz-Mach-Poincar{\'e} (BLMP).
 Let the reference body of  the first, 
inertial frame (IRF), 
is associated with the surface of a non-rotating planet, and the reference body of 
the second frame formed by a set of material points, falling freely under the 
influence of the planet gravity.
 (It can be named by proper reference frame of the given force field (PRF)).
	
The observer, located in the first, inertial, frame of reference, of course will 
examine the fall of test bodies as happens under the action of a force field $\mathcal{F}$
  in 
the Minkowski space-time, the source of which is the planet. 
He sees no need to 
explain the motion of test bodies with curvature. 

However, the observer, located in
 the second reference frame, 
 does not detect this force field. 
Instead,  he observes rapprochement of points of the reference body 
of his frame which for him  are points of his  physical space.  

 If he is denied the opportunity to see the planets and stars, 
it seems impossible for him to find another explanation of this fact, which is
different from the generally accepted explanation --- of an evidence of space-time curvature.

Thus, if an observer in a IRF can consider
space-time as flat, then the observer in the PRF 
 of the force field $\mathcal{F}$, who proceeds from  relativity of
 space and time in the BLMP meaning, \textit{is forced} to
consider it as an non-Euclidean.

Some quantitative results on the metric of space-time in PRFs were obtained earlier by 
the author \cite{Verozub08a}. 
 Namely, we postulate that space-time $E$
in inertial frames is 
the Minkowski  one, according  to the spacial relativity. 
 From our point of view, space-time geometry and properties of the reference frame 
do not 
have meaning by themselves. Therefore, this postulate means that only a 
complex ``Minkowski 
space-time $E$ + inertial reference system'' makes a physical sense. 
Starting from this 
postulate and based
 on the relativity of space-time, it is possible to find the line element of 
space-time $V$ in a
PRF  of any given in the $E$ force field.

Consider a PRF,  the reference body  of which  formed by  material points with 
masses $m$  moving under the action of the force field $\mathcal{F}$. 
If we proceed from relativity of space and time in the BLMP
 sense, then
the line element of space-time in PRFs can be expected to have
the following form \cite{Verozub08a}
\begin{equation}
ds=-(mc)^{-1}\,dS(x,dx). \label{dsMain},%
\end{equation}
where
$dS=\mathcal{
L}(x,\dot{x})dt$, and $\mathcal{L}(x,\dot{x})$ is a Lagrange function
describing in Minkowski space-time the motion of the identical point masses $m$.

\section{Examples}

1. Suppose that in the Minkowski space-time gravitation can be described
as a tensor field  $\psi_{\alpha\beta}(x)$ in $E$, and the Lagrangian,
describing the motion of a test particle with the mass $m$ in $E$  is given by
the form
\begin{equation}
\mathcal{L}=-m c [g_{\alpha\beta}(\psi)\;\dot{x}^{\alpha}\;\dot
{x}^{\beta}]^{1/2},  \label{LagrangianThirr}
\end{equation}
where $\dot{x}^{\alpha}=dx^{\alpha}/dt$ and $g_{\alpha\beta}$ is a
symmetric tensor whose components are 
functions of $ \psi_{\alpha\beta} $  \cite{Thirring}.

If particles move under influence of the force field $\psi_{\alpha\beta}(x)$,  
then according to (\ref{dsMain})
 the space-time line element in PFRs of this field takes the form
\begin{equation}
ds^{2}=g_{\alpha\beta}(\psi)\;dx^{\alpha}\;dx^{\beta}
\end{equation}
Consequently,  the space-time in
such PRFs is  Riemannian  $V$  with  curvature other than zero.
The tensor $g_{\alpha\beta}(\psi)$  is a  space-time metric tensor in  the PRFs. 


Viewed by an
observer located in the IRF, the motion of the particles, forming the
reference body of the PRF, is affected by the force field $\psi_{\alpha\beta} $. 
Let $x^{i}(t,\chi)$ be a set of the particles paths, depending on the parameter $\chi$.
 Then, for the observer located in the IRF the
relative motion of a pair of particles from the set is described in
non-relativistic limit by the differential equations \cite{MTU} 
\begin{equation}
\frac{\partial^{2}n^{i}}{\partial t^{2}}+\frac{\partial^{2}U}{\partial
x^{i}\partial x^{k}}n^{k}=0,  \label{DevIFR}
\end{equation}
where $n^{k}=\partial x^{k}/\partial{\chi}$ and $U$ is the gravitational
potential.

However, the observer in a PRF of this field will not feel the existence of
the field.
The presence of the field $\psi_{\alpha\beta}$ will be displayed for him
differently --- as space-time curvature which manifests itself as a deviation
of the world lines of nearby points of the reference body.

For a quantitative description of this fact it is natural for him to use the
Riemannian normal coordinates.
\footnote{This and the above consideration does not depend on the used coordinate 
system, it can be performed by a covariant method.}
 In these coordinates spatial components of
the deviation equations of geodesic lines are 
\begin{equation}
\frac{\partial n^{i}}{\partial t^{2}}+R_{0k0}^{i}n^{k}=0,
\end{equation}
where $R_{0k0}^{i}$ are the components of the Riemann tensor. 
In the Newtonian limit these equations coincide with (\ref{DevIFR}).

Thus, in two  frames of reference being used we have two different descriptions of 
particles 
motion --- as moving under the action of a force field in the Mankowski space-time, and as 
moving 
along the geodesic line in a Riemann space-time
with the curvature other than zero.

2. Another, rather unexpected example, give the recent results on the motion of small
elements of a perfect isentropic fluid \cite{Verozub08a}.

Instead of the traditional continuum assumption, the behavior of the fluid flow can 
be considered as the motion of a finite mumber of particles uder the influence of interparticles
 forces which mimic effects of pressure, viscosity, etc. \cite{Monaghan}. 
Owing to replacement of integration by summation over a number of particles,   continual
 derivatives  become simply  time derivatives along  the particles trajectories. 
The velocity of the fluid at a given point is the velocity of the particle at this point. 
The continuity equation is always fulfilled and can consequently be omitted.  
Owing to such discratization the motion of particles is governed by means of solutions of 
ordinary differential equations of classical or relativistic dynamics.

In \cite{Verozub08a} it was shown  
 that the following Lagrangian describes the motion of small elements of a perfect isentropic 
fluid in adiabatic processes is given by
\begin{equation}
L=-mc \left( G_{\alpha \beta }\frac{dx^{\alpha }}{d\lambda }\frac{%
dx^{\beta }}{d\lambda }\right) ^{1/2}d\lambda   \label{Lagrangian_in_V}.
\end{equation}%
In this equation 
 $G_{\alpha \beta }=\varkappa ^{2}\eta_{\alpha \beta }$,%
 $\eta_{\alpha \beta}$ is the metric tensor of the  space-time $E$,

\begin{equation}
\varkappa =\frac{w}{nmc^{2}}=1+\frac{\varepsilon }{\rho c^{2}}+\frac{P}{\rho
c^{2}},  \label{xi}
\end{equation}%
$\varepsilon $ is the fluid density energy, $m$ is the mass of the fluid particles,  
$c$ is speed of light, and $\rho=m n$, $n$ is the particles number density, $P$ is the 
pressure i the fluid,
$\lambda $ is a parameter along  4-pathes of particles.

In an inertial reference drame (i.e. in Minkowski space-time $E$)
 we can set the parameter $\lambda =\sigma $
which yields the following  Lagrange equations:

\begin{equation}
\frac{d}{d\sigma }\left( \varkappa u_{\alpha }\right) -\frac{\partial
\varkappa }{\partial x^{\alpha }}=0
\end{equation}%
where $u_{\alpha }=\eta _{\alpha \beta }u^{\beta }$, and $u^{\alpha}=dx^{\alpha}/d\sigma$. 
 For adiabatic processes 
\cite{Landau} 
\begin{equation}
\frac{\partial }{\partial x^{\alpha}}\left( \frac{w }{n}\right) =\frac{1}{n}\frac{\partial P%
}{\partial x^{\alpha }},
\end{equation}%
and we arrive at the equations of the motion of the set of the particles in the form%
\begin{equation}
w\frac{du_{\alpha }}{d\sigma }+u_{\alpha }u^{\beta }\frac{\partial P}{%
\partial x^{\beta }}-\frac{\partial P}{\partial x^{\alpha }}=0.
\label{MotionEquation_in_E}
\end{equation}%
where $du_{\alpha }/d\sigma =\left( \partial u_{\alpha }/\partial
x^{\epsilon }\right) u^{\epsilon }.$ It is the general accepted relativistic
equations of the motion of fluid  \cite{Landau}.

In a comoving reference frame the space-time the line element is of the form
\begin{equation}
ds^{2}=G_{\alpha \beta }dx^{\alpha }dx^{\beta }  \label{ds2}.
\end{equation}%

In this case the element of the proper time is $ds$. After the seting 
$\lambda =s$,
the Lagrangian equation of the motion takes the standard form of a congruence of 
geodesic lines :%
\begin{equation}
\frac{du^{\alpha }}{ds}+\Gamma _{\beta \gamma }^{\alpha }u^{\beta }u^{\gamma
}=0,  \label{eqsPart_as_Geodesic}
\end{equation}%
where $du_{\alpha }/ds=\left( \partial u_{\alpha }/\partial x^{\epsilon
}\right) u^{\epsilon }$, $u^{\alpha}=du^{\alpha}/ds$,
and 
\begin{equation}
\Gamma _{\beta \gamma }^{\alpha } =\frac{1}{2}G^{\alpha \epsilon }\left( 
\frac{\partial G_{\epsilon \beta }}{\partial x^{\gamma }}+\frac{\partial
G_{\epsilon \gamma }}{\partial x^{\beta }}-\frac{\partial G_{\beta \gamma }}
{\partial x^{\epsilon }}\right).
\end{equation}
In the Cartesian coordinates
\begin{equation}
\Gamma_{\alpha\beta}^{\gamma}=
\frac{1}{\varkappa }\left( \frac{\partial \varkappa }{\partial x^{\gamma }%
}\delta _{\beta }^{\alpha }+\frac{\partial \varkappa }{\partial x^{\beta }}%
\delta _{\gamma }^{\alpha }-\eta ^{\alpha \epsilon }\frac{\partial \varkappa 
}{\partial x^{\epsilon }}\eta _{\beta \gamma }\right),
\end{equation}
so that 
\begin{equation}
 \Gamma_{00}^{1}= -\frac{1}{\rho c^{2}}\frac{\partial P}{\partial x^{1}}
\end{equation}

In the spherical coordinates the scalar curvarure $R$ is given by
\begin{equation}
 R = \frac{6}{\varkappa^{3} r^{2}} (r^{2} \varkappa')', 
\end{equation}
where the prime denotes a derivative with respect to $r$.

Therefore, the motion of small elements of the fluid in a comoving reference 
frame can be viewed as the motion in a Riemannian space-time with a nonzero 
curvature.

Of course,  (\ref{dsMain}) refers to any classical field $\mathcal{F}$. For instance,
space-time in PRFs of an electromagnetic field is Finslerian. However, since 
$ds$, in this case, depends on the mass and charge  of the particles
forming the reference body, this fact is not of great significance.

Thus any force field  can be considered based on 
 the aggregate  "IRF + Minkowski space", and based on  the  aggregate 
 "PRF + non-Euclidean space-time with metric (\ref{dsMain})"
From this point of view of  geometrization of gravity is the second possibility, 
which was discovered by Einstein's intuition.

It is important to realize that the relativity of space-time geometry to the 
frame of reference is the same important and fundamental property of physical relativity 
as relativity to  act of measurement, the physical 
realization of which is quantum mechanics. Full implementation of these ideas 
can have far-reaching implications for fundametal physics.


\section{Gravity equations and gauge-invariance}
In the theory of gravitation, the equations of motion of test particles play a fundamental 
role.
 Notion of "gravitational field" emerged as something necessary to correctly describe the 
motion of bodies. The values that appear in the equations of motion, 
become the main characteristic of the field. The field equations have emerged as a tool for 
finding these values for a given  distribution of masses.

All this is very similar to
 classical electrodynamics. It is  very important in this case 
that the equations of motion of
 test charges are invariant under  gauge transformations of 4-potentials. For this reason, 
all  4-potentials, obtained from a given by a gauge transformation, describe the same 
field.
 That is why the field equations of classical electrodynamics are invariant under gauge 
transformations.

Einstein's  equations of the motions of test particles in gravitational field 
are also invariant with respect to some class of transormations of the field 
variables in  any given coordinate system  --- 
with respect to geodesic transformations of  Christoffel symbols (or metric tensor).
\cite{Eisenhart}
 Such transformations for 
 the Christoffel symbols are of the form
\begin{equation}
\label{GammaGeodesTransformations}\overline{\Gamma}_{\beta\gamma}^{\alpha
}(x)=\Gamma_{\beta\gamma}^{\alpha}(x)+\delta_{\beta}^{\alpha}\ \phi_{\gamma
}(x)+\delta_{\gamma}^{\alpha} \phi_{\beta}(x),
\end{equation}
where $\phi_{\alpha}(x)$ is  a continuously differentiable vector
field.
(The transformations for the metric tensor are solutions of some complicate 
partial differential equations).

Consequently, all  Christoffel symbols obtained from a given by geodesic 
transformations,
 describe the same gravitational field. 
 The equations for determining the gravitational field must be invariant under 
such 
transformations, and the physical meaning can only have values which are invariant under
 geodesic transformations.

However, Einstein's gravitational equations are not consistent completely
with the requirements 
which imposes on them the main hypothesis of the motion of test particles along geodesics,
 because they are not geodesically invariant \cite{Petrov}. 

Therefore, we can assume that in a fully correct theory of gravity, based
 on the hypothesis 
of the motion of test particles along geodesics, geodesic transformations should play the 
role 
of gauge transformations, and coordinate transformation should play the same role as in 
electrodynamics.

Einstein equations are in good agreement with observations in weak and moderately strong 
fields.
Therefore, if there are more correct equation of gravitation, then deriving from them  
physical results should differ  observably from   Einstein's equations only in strong fields. 

Simplest vacuum equation of this kind 
 were first proposed (from a different point of view) in \cite{Verozub91}, 
and discussed in greater detail in \cite{Verozub08a},
  their physical implications discussed in 
\cite{Verozub96} - \cite{Verozub06}, 
and the equations in the presence of matter - in \cite{VerKoch00}. 
They are some  geodesic-invariant modification of Einstein's equations.

From a theoretical point of view, the most satisfactory are the vacuum equations. 

They predict some fundamentally new physical consequences which can be tested 
experimentally.


Under geodesic transformations the Ricci tensor $R_{\alpha\beta}$ of space-time $V$  
in PRFs of gravitational field transforms 
as follows:
\begin{equation}
\overline{R}_{\alpha\beta}=R_{\alpha\beta}+(n-1)\psi_{\alpha\beta},
\end{equation}
where
\begin{equation}
\psi_{\alpha\beta}=\psi_{\alpha;\beta}-\psi_{\alpha}\psi_{\beta,}
\end{equation}
 and a semicolon denotes a covariant differetiation 
in $V$. Therefore, the simpest generalization of the Einstein
equations is of the form
\begin{equation}
R_{\alpha\beta}+(n-1)\Gamma_{\alpha\beta}=0,
\end{equation}
 where $\Gamma_{\alpha\beta}$ is a tesor transformed under geodesic
transformations as follows
\begin{equation}
\overline{\Gamma}_{\alpha\beta}=\Gamma_{\alpha\beta}-\psi_{\alpha\beta}.
\end{equation}

Due to the fact that our space-time is a bimetric, there exists a
vector field
\begin{equation}
Q_{\alpha}=\Gamma_{\alpha}-\overset{\circ}{\Gamma}_{\alpha}
\end{equation}
where $\Gamma_{\alpha}=\Gamma_{\alpha\beta}^{\beta}$ , $\overset{\circ}{\Gamma}_{\alpha}=
\overset{\circ}{\Gamma}_{\alpha\beta}^{\beta}$
,  $\Gamma_{\alpha\beta}^{\gamma}$ and 
$\overset{\circ}{\Gamma}_{\alpha\beta}^{\gamma}$
 are the Christoffel symbols  
in $V$ and $E$, respectively.

Under geodesic transformations in $V$  the quantities $\Gamma_{\alpha}$ are transformed 
as follows:

\begin{equation}
 \overline{\Gamma}_{\alpha} =\Gamma_{\alpha} + (n+1)\,\psi_{\alpha}
\end{equation}
For this reason, a tensor object 
\begin{equation}
 A_{\alpha\beta}=Q_{\alpha;\beta}-Q_{\alpha}Q_{\beta},
\end{equation}
 where $Q_{\alpha;\beta}$ is a covariant derivative of $Q_{\alpha}$ in $V$, has the 
same transformation properties under geodesic transformations as  must have the above 
vector field $\Gamma_{\alpha\beta}$.

The line element of space-time in PRFs was obtained from 
the Lagrangian motion of test particles in the Minkowski space-time $E$. 
If we want to find the equation of gravity in space-time $E$, you must realize that
 in this space, 
the Christoffel symbols $\Gamma_{\alpha\beta}^{\gamma}$ 
can be regarded as components of the tensor $\Gamma_{\alpha\beta}^{\gamma}-
\overset{\circ}{\Gamma}_{\alpha\beta}^{\gamma}$ in the Cartesian coordinate system,
 i.e.  
as components of $\Gamma_{\alpha\beta}^{\gamma}$, where the ordinary derivatives 
replaced by covariant in the metric of space-time $E$.
(Just as in bimetric Rosen's theory  \cite{Treder}). 

Given this, we arrive at the conclusion that the equation
\begin{equation}
 R_{\alpha\beta} - A_{\alpha\beta}=0
\end{equation}
is the simplest geodesic invariant modification of the vacuum Einstein equations, considered 
from the point of view of flat space-time.

These equations can be written in another form.
The simplest geodesic-invariant object in $V$ is a Thomas symbols:
\begin{equation}
 \Pi_{\alpha\beta}^{\gamma}=\Gamma_{\alpha\beta}^{\gamma}-
\frac{1}{n+1}
\left(\delta_{\alpha}^{\eta} \Gamma_{\beta}+ 
      \delta_{\beta}^{\eta} \Gamma_{\alpha} \right). 
\end{equation}
It is  not a tensor. However, from point of view of flat space-time $E$, 
they can be considered as components of the tensor $B_{\alpha\beta}^{\gamma} =
\Pi_{\alpha\beta}^{\gamma}-\overset{\circ}{\Pi}_{\alpha\beta}^{\gamma}$,
where $\overset{\circ}{\Pi}_{\alpha\beta}^{\gamma}$ is the Thomas simbols in $E$.
In another words, $B_{\alpha\beta}^{\gamma}$ can  
be considered  as the Thomas symbols where  derivatives
replaced by the covariant ones with respect to the metric $\eta_{\alpha\beta}$.
This geodesic-invariant tensor can be named by strength tensor of gravitational field.

The above gravitation equation can be written by tensor $B_{\alpha\beta}^{\gamma}$ 
as follows:
\begin{equation}
\label{myequations}
\bigtriangledown_{\gamma} B_{\alpha\beta}^{\gamma}-
B_{\alpha\delta}^{\gamma} B_{\beta\gamma}^{\delta}=0.
\end{equation}
where $\bigtriangledown$ denotes  a covariant derivative in $E$.

The physical consequences following from these equations 
 do not contradict any observational data, however, lead to some unexpected results, 
which allow to us to test the theory.
The first result is that they predict the existence of supermassive compact objects 
without event horizon which are an alternative to supermassive black holes in the centers 
of galaxies.
 The second result  is that they provide a simple and natural explanation for the 
fact of an acceleration of the universe as of a consequence of the gravity properties.

\section{Remaks on the equations inside matter}

 We can not claim that the particles inside the material medium move along geodesics. 
The exception is the case of dust matter and perfect fluid. 
Consequently,  it is unclear whether the field equations inside the matter
 to be a generalization of the geodesic equations of Einstein. 
However, such equations have been proposed in the work \cite{VerKoch00}. 
Comparison of the results obtained from 
them with observations of the binary pulsar PSR 1913+16 shows good agreement with 
observations. 
Despite this,  doubts   as to their correctness are still remain.
The problem is that the writing of 
generalization of the  equations in the matter requires 
significantly 
narrow the class of admissible geodesic transformations of the metric tensor of 
space-time $V$. It is not clear  whether such space-time is Riemannian.
It is possible,  geodesic invariance is violated in a material medium.
For this reason,  we do not consider these equations here in 
more detail, assuming that this is still a subject for further research.

\end{document}